\documentclass[twocolumn,secnumarabic,amssymb, aps,prl,reprint]{revtex4-1}
\usepackage{graphicx,siunitx}
\usepackage{dcolumn}
\usepackage{bm}
\usepackage{ucs}

\usepackage{mathtools}
\begin{document}

\title{Supercurrent through a spin-split quasi-ballistic point contact in an InAs nanowire}

\author{J. C. Estrada Salda\~{n}a$^{1}$}
\thanks{Present address: Center for Quantum Devices, Niels Bohr Institute, University of Copenhagen, Copenhagen, Denmark}
\author{R. \v{Z}itko$^{2}$}
\author{J. P. Cleuziou$^{1}$}%
\author{E. J. H. Lee$^{1}$}%
\thanks{Present address: Condensed Matter Physics Center (IFIMAC), Universidad Aut\'{o}noma de Madrid, 28049 Madrid, Spain.}
\author{V. Zannier$^{3}$}
\author{D. Ercolani$^{3}$}
\author{L. Sorba$^{3}$}
\author{R. Aguado$^{4}$}
\author{S. De Franceschi$^{1}$}
\email{Corresponding author: silvano.defranceschi@cea.fr}

\affiliation{$^{1}$Univ. Grenoble Alpes, INAC-PHELIQS, F-38000 Grenoble, France and CEA, INAC-PHELIQS, F-38000 Grenoble, France}
\affiliation{$^{2}$Jo\v{z}ef Stefan Institute, Jamova 39, Ljubljana, Slovenia \\
Faculty of Mathematics and Physics, University of Ljubljana, Jadranska 19, Ljubljana, Slovenia}
\affiliation{$^{3}$ NEST-Istituto Nanoscienze-CNR and Scuola Normale Superiore, Piazza S. Silvestro 12, 56127 Pisa, Italy}
\affiliation{$^{4}$Materials Science Factory, Instituto de Ciencia de Materiales de Madrid (ICMM), Consejo Superior de Investigaciones Cient\'{i}ficas (CSIC), Sor Juana In\'{e}s de la Cruz 3, 28049 Madrid, Spain}

\date{\today}

\begin{abstract}

We study the superconducting proximity effect in an InAs nanowire contacted by Ta-based superconducting electrodes. Using local bottom gates, we control the potential landscape along the nanowire, tuning its conductance to a quasi-ballistic regime. At high magnetic field ($B$), we observe approximately quantized conductance plateaus associated with the first two spin-polarized one-dimensional modes. For $B < 1$ T, the onset of superconductivity occurs in concomitance with the development of sizeable charge localization leading to a 0.7-type conductance anomaly. In this regime, the proximity supercurrent exhibits an unusual, non-monotonic $B$ dependence. We interpret this finding in terms of a competition between the Kondo effect, dominating near $B=0$, and the Zeeman effect, enforcing spin polarization and the emergence of a $\pi$ phase shift in the Josephson relation at higher $B$.

\end{abstract}

\maketitle

One-dimensional (1D) semiconductor nanowires (NWs) with strong spin-orbit coupling and induced superconductivity are attracting considerable attention owing to their potential to realize topological superconductivity and emergent Majorana modes \cite{lutchyn2010majorana,Oreg2010,lutchyn2017review,Aguado2017review}. For a topological phase to be established, the 1D character has to be preserved over micron-scale lengths and the chemical potential needs to be positioned within the so-called helical gap opened by a properly oriented magnetic field $B$ \cite{chen2017experimental}. Given the modest size of the spin-orbit energy \footnote{of the order of $\sim 100 \mu$eV according to tight-binding atomistic and self-consistent simulation of the band structure of an InAs nanowire of diameter = 20 nm (Zaiping Zeng and Yann-Michel Niquet, private communication). See also, Ref. \cite{nadj2012spectroscopy}.}, the second condition implies that the 1D conduction mode supporting Majoranas should be only slightly filled. For this reason, it is important to explore the 1D properties of semiconductor NWs at low subband filling, in the presence of  the superconducting proximity effect and an externally applied magnetic field. To this aim, we investigate InAs NWs in combination with tantalum-based superconducting contacts with a high in-plane critical magnetic field, $B_c \sim  1.8 T$.
 
Conductance quantization is the most commonly observed experimental signature of ballistic 1D transport \cite{VanWees1988}. In semiconductor NWs, this phenomenon is more easily observed at large magnetic field \cite{doi:10.1021/nl3035256}, where backscattering is reduced and spin degeneracy is simultaneously lifted, leading to conductance steps of $e^2/h$, where $e$ is the electron charge and $h$ the Planck constant.  More recently, conductance quantization was observed also at zero magnetic field, with steps of  $2e^2/h$ due to two-fold spin degeneracy \cite{heedt2016ballistic,Kammhuber2016,zhang2016ballistic,fadaly2017observation,gooth2017ballistic,Estrada2017}.
Here we make use of two independently tunable bottom gates in order to tailor the potential landscape in the NW channel \cite{Estrada2017}. Proper tuning of the applied gate voltages results in the creation of a local point contact exhibiting approximately quantized conductance plateaus in the few-channel regime. Interestingly, we find that unintentional charge localization, while seemingly suppressed at high magnetic field, becomes apparent at low field, resulting in a 0.7-type conductance anomaly, a phenomenon largely studied in quantum point contacts formed within high-mobility two-dimensional heterostructures \cite{PhysRevLett.88.226805,thomas1996possible,sfigakis2008kondo,iqbal2013odd} and only recently in NWs \cite{heedt2016ballistic}. 

In this exotic regime, and thanks to the large electron g-factor in InAs and to the relatively large $B_c$, we are able to investigate the superconducting proximity effect coexisting with a strong $B$-field-induced spin polarization. We observe a non-monotonic behavior of the critical current as a function of $B$ field that can be understood as a Zeeman-driven quantum phase transition from a spin singlet ground state, with 0-phase-shift Josephson coupling, to a spin-1/2 ground state, with $\pi$-phase-shift Josephson coupling. Upon increasing the magnetic field, the supercurrent first vanishes at the $0$-$\pi$ transition, and then recovers once the Zeeman energy is large enough to stabilize the spin-1/2 ground state, resulting in a non-monotonic $B$ dependence. This interpretation is confirmed by theoretical calculations based on an Anderson-type model coupled to superconducting leads with strong and gate-dependent tunnel couplings. 

\begin{figure}
\includegraphics[width=86mm]{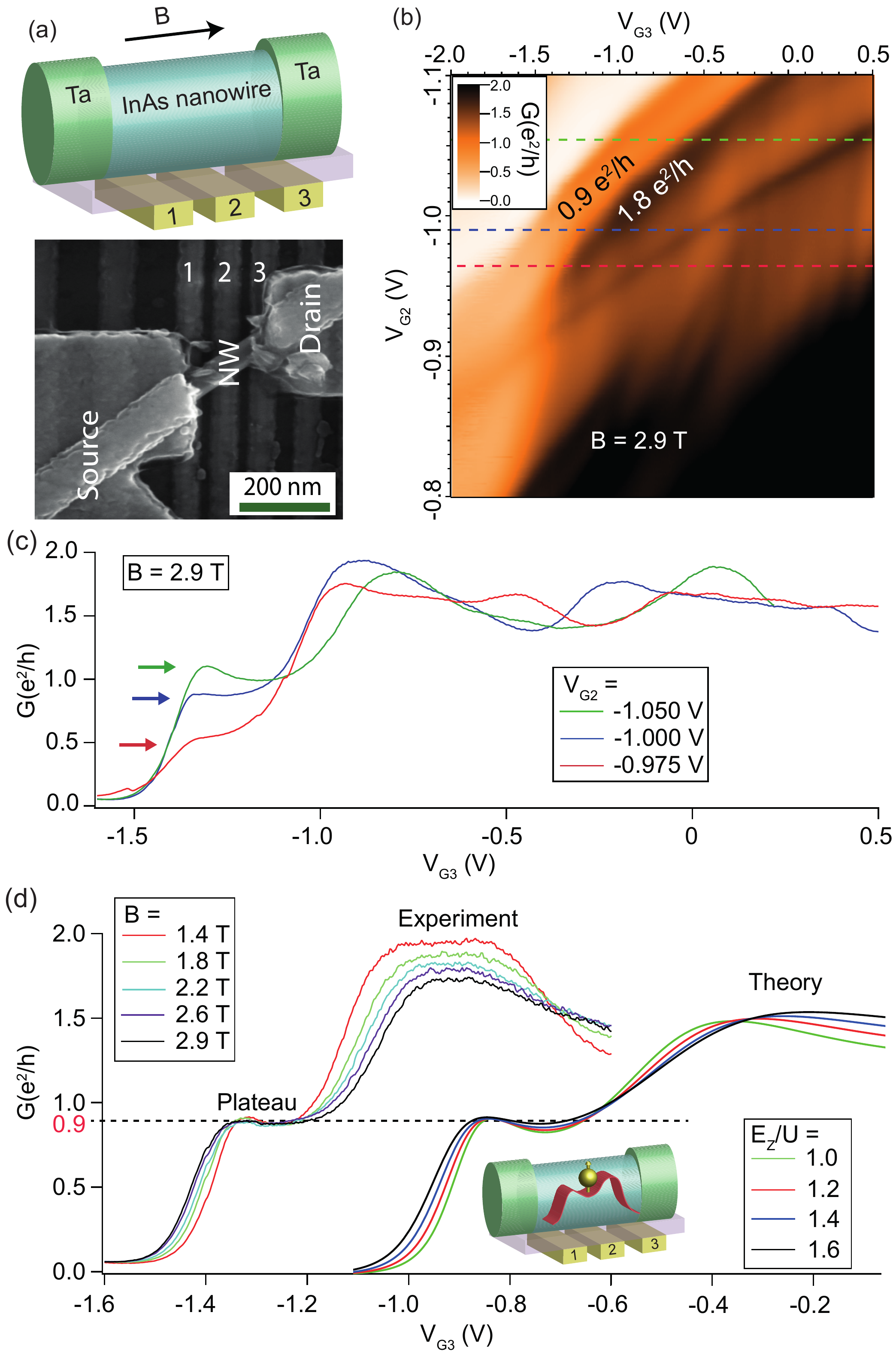}
\caption{(a) Schematics and scanning-electron micrograph of the device. (b) Normal state ($B$ = 2.9T) measurement of the linear conductance, $G = dI/dV(V_{sd}=0)$, as a function of $V_{G2}$ and $V_{G3}$. Near pinch-off, two conductance plateaus appear at $G \approx 0.9 e^2/h$ and $1.8e^2/h$. 
(c) $G(V_{G3})$ curves taken at $V_{G2}$ = -0.975, -1 and -1.05 V, as indicated by the dashed lines in panel (b). (d) Left: $G(V_{G3})$ curves measured at different $B$ ($V_{G2}$ = -1V). The conductance of the $0.9e^2/h$ plateau remains unchanged within the explored $B$ range. Right: NRG simulations of $G(V_{G3})$ at different values of the Zeeman energy, $E_z$ normalized to the charging energy $U$. The experimental and theory curves are shifted horizontally for clarity. Inset: schematic representation of a camel-shape, conduction-band profile created by the local gates and the associated charge localization.}
\label{fig:image1}
\end{figure}

The device designed for our experiment is shown in Fig. \ref{fig:image1}a. It was fabricated from a single, 65-nm-diameter InAs NW grown by chemical beam epitaxy \cite{gomes2015controlling}. 
The NW was deposited on a bed of narrow gate electrodes covered by 12 nm of HfO$_2$. Successively, Ta (60 nm)/Al (15 nm) source and drain contacts with a spacing of 280 nm were defined by e-beam lithography and subsequent e-beam evaporation. The latter was preceded by a gentle in-situ Ar etching to remove the native oxide of the NW. The Ta/Al contacts were measured to be superconducting below a critical temperature, $T_c \sim 0.8$ K, which is consistent with values reported for Ta in the crystalline $\beta$-phase ($T_c=0.67-0.9$ K \cite{schwartz1972temperature}). 
The sample was mounted in a dilution refrigerator (base temperature of 15 mK), and a magnetic field, $B$, was applied in the device plane using a 
vector magnet. In all of the measurements presented here, $B$ was aligned with the longitudinal axis of the NW (data for different angles can be found in the Supplementary Material). 

In order to look for conductance quantization, the device was first brought to the normal state by applying a high magnetic field B = 2.9 T, i.e. well above $B_c$; the linear conductance, $G = dI/dV(V = 0)$, was measured as a function of voltages $V_{G2}$ and $V_{G3}$, applied to gates 2 and 3, respectively (Fig. \ref{fig:image1}b). Two conductance  plateaus, around  $0.9 e^2/h$  and $1.8 e^{2}/h$, can be identified,  i.e. close to  the ideally expected values for one and two 1D modes $e^{2}/h$ and $2e^{2}/h$,  respectively. As we shall see further below, these modes are resulting from the Zeeman-induced splitting of the first spin-degenerate 1D subband. 

To a closer look, Fig. \ref{fig:image1}b shows noticeable structures consisting of conductance modulations of up to 20\% superimposed on the quantized plateaus. We ascribe these modulations to tunneling resonances associated mainly with quasi-localized states in the quantum point contact. Such states are expected to have similar capacitive coupling to gates $G$2 and $G$3, hence producing the predominantly diagonal conductance ridges observed in Fig. \ref{fig:image1}b. The amplitude of these  additional features varies over the ($V_{G2}$,$V_{G3}$) plane and can vanish at certain regions.

Figure \ref{fig:image1}c shows three $G(V_{G3})$ traces taken at different $V_{G2}$ as indicated by the horizontal dashed lines in Fig. \ref{fig:image1}b.
The green trace exhibits a clearly visible broad peak structure causing an overshoot of the conductance at the onset of the first plateau. This structure is no longer present in the blue trace resulting in an essentially flat conductance plateau.
Further increasing $V_{G2}$ results in a global suppression of the conductance step (red trace).

From now on we focus our attention on the intermediate value of $V_{G2}$, where the first conductance step is ``cured'' from spurious resonances thereby resembling the one expected for the onset of the first 1D conduction mode in a ballistic point contact. From a comparison with the other traces we know that a resonance is in fact lurking in this seemingly ideal conductance plateau. This underlines the importance of double gate control in revealing the nature of the observed transport features. The underlying presence of charge localization is on the other hand apparent in the second plateau where  conductance oscillations remain clearly visible (Fig. \ref{fig:image1}c \footnote{We note that the second plateau extends on a much larger $V_{G3}$ range suggesting that conductance is limited by the barrier induced by $V_{G2}$.}).

\begin{figure}[!htb]
\includegraphics[width=86mm]{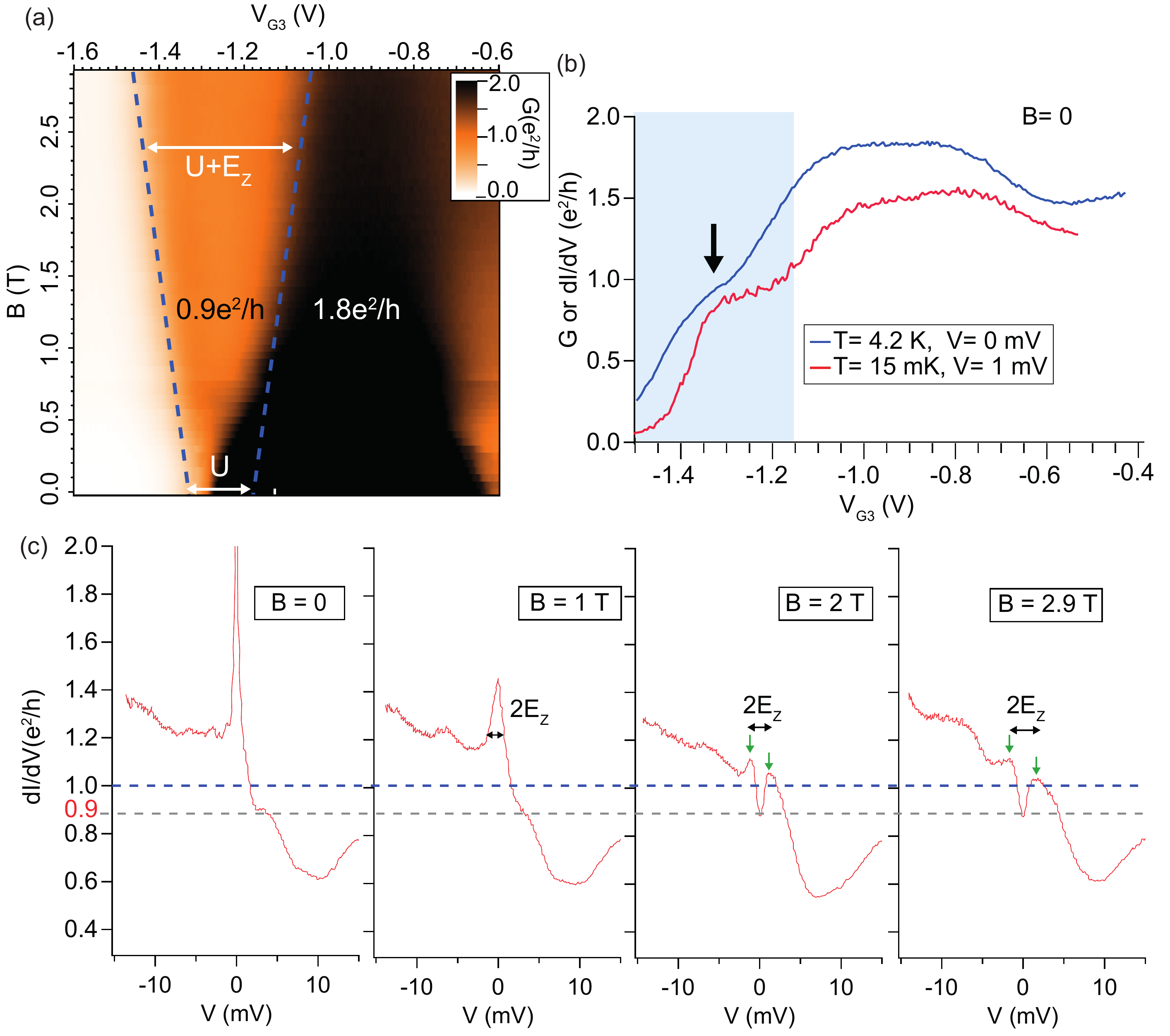}
\caption{(a) $G$ as a function of $B$ and $V_{G3}$. The Zeeman-split conductance plateaus at higher $B$ (onset marked by dashed lines) display a residual zero-field splitting, indicative of a localized charge state with
charging energy $U$. (b) Normal state conductance as a function of $V_{G3}$ at $B=0$. Blue trace:  $G(V_{G3})$ measured well above the superconducting critical temperature. Red trace: differential conductance, $dI/dV$, measured at T = 15 mK and $V =1$ mV, i.e. well above the superconducting gap. Both curves display a shoulder preceding the $2 e^2/h$ conductance plateau (black arrow), characteristic of the 0.7 anomaly in QPCs. The blue-shaded region highlights the $V_{G3}$ range in which the supercurrent in Figure \ref{fig:image3} is studied.
(c) $dI/dV(V)$ measurements at different $B$ (T = 15 mK). $V_{G3}$ is fixed at the position of the 0.7 anomaly. The $B$-field evolution of the observed zero-bias peak  is consistent with the Zeeman splitting of a Kondo resonance.The divergent zero-bias peak at $B=0$ is due to a supercurrent.  
}
\label{fig:image2}
\end{figure}

The first conductance plateau preserves its flat, featureless character over a relatively large magnetic field range. Upon reducing $B$ down to 1.4 T (Fig. \ref{fig:image1}d), the plateau shrinks with $B$ due to the decreasing Zeeman energy, $E_Z = g \mu_B B$, with $\mu_B$ being the Bohr magneton and $g$ the electron g-factor in the point contact, while the conductance remains quantized at  $0.9 e^{2}/h$. Below 1.4 T, a conductance enhancement begins to emerge due to supercurrents (not shown). The full-range B-field dependence is shown as a color-scale plot in Fig. \ref{fig:image2}a.
 
The superimposed dashed lines highlight the $B$-evolution of the first conductance plateau. Interestingly, the two lines do not coalesce at $B=0$ as we may expect if the width of the plateau were proportional to $E_Z$.  Instead, a residual zero-field splitting remains. Its origin can be ascribed once again to a localized charge state, most likely the evolution to zero field of the one already identified at $B=2.9$ T. The residual splitting is indicative of a sizable charging energy ($U\sim1.3$ meV with $|g| = 11$ \footnote{Converting the $V_{G3}$ scale into energy is possible with the help of differential conductance ($dI/dV$) measurements at finite source-drain bias voltage, $V_{sd}$. This standard procedure (Supplementary Material) yields a conversion factor $\alpha$ = 0.0082 meV/mV}) associated with the localized state. 

Localized states are often observed in semiconductor NWs. They can form \cite{weis2014quantum,schroer2010correlating,voisin2014few} due to a plethora of confining mechanisms: crystal defects or impurities in the NW, tunnel barriers at the contacts, surface charges, 
or Friedel oscillations in electron density. In a gate-defined point contact, where charge density is substantially lowered and electric-field screening consequently reduced, localization is enhanced and Coulomb interaction emerges. Strongly localized states leading to a few rather sharp Coulomb resonances can indeed be observed in the studied gate-induced constriction near full charge depletion. They lie at further negative gate voltages, outside the ($V_{G2}$, $V_{G3}$) field explored in Fig. \ref{fig:image1}b (see Supplementary Material). The localized state at the onset of the first conductance plateau has a more subtle nature, and, as we have seen, its presence may go unperceived without a proper control of the electrostatic landscape.

At $B=0$, transport is largely affected by the superconducting proximity effect. A dissipation-less supercurrent sets in already at the onset of the first quasi-ballistic conduction mode leading to a divergence of the conductance. Before discussing the superconducting regime it is instructive to examine the normal type behavior, which can be accessed at temperatures above $T_c$. Figure \ref{fig:image2}b shows a characteristic $G(V_{G3})$ measurement at 4.2 K. Interestingly, the onset of conduction through the first spin-degenerate subband (whose conductance is around 1.5 $e^2/h$, i.e. somewhat lower than at B = 2.9 T) is preceded by a shoulder at $\sim e^2/h$. A shoulder can also be consistently found in a measurement of $dI/dV$ at
$eV_{sd} \gg \Delta$ and $T$ = 15 mK, where $\Delta$ is the superconducting gap.

This feature is reminiscent of a phenomenon known as the "0.7 anomaly'', largely studied in conventional quantum point contacts \cite{VanWees1988} electrostatically defined in a high-mobility two-dimensional electron gas . The origin of the 0.7 anomaly has received a variety of explanations raising a long-standing debate \cite{micolich2011lurks}. A number of works point toward Kondo-effect physics associated with a localized state in the QPC \cite{PhysRevLett.88.226805,Heyder2015,Brun2016,Meir2002}, a physical picture that, in view of the already discussed observations, appears be appropriate to the system studied here. This picture is further confirmed by $dI/dV(V_{sd})$ measurements for different values of $B$ and fixed $V_{G3} = -1.25 V $ at the $e^2/h$ shoulder. The data, plotted in Fig. \ref{fig:image2}c, show the characteristic Zeeman splitting of a zero-bias Kondo resonance. At $B=0$, the resonance has a zero-bias divergence due to the superconducting proximity effect.

To confirm our interpretation in terms of a quasi-ballistic QPC with a 0.7-type anomaly arising from a localized spin-1/2 state,
we model the device using the following Anderson-type Hamiltonian:

\begin{equation}
\label{hamiltonian}
\begin{split}
H&=\delta (n-1) + U/2 (n-1)^2 + E_z S_z + \sum_{k\sigma} \epsilon_k c^\dag_{k\sigma}
c_{k\sigma} + \\
&+ \sum_{k\sigma} \left(v(1-x n_{d,\bar{\sigma}}) c^\dag_{k\sigma} d_\sigma +  \text{H.c.} \right)  + \\
&+ \Delta \sum_{k} (c_{k\uparrow}c_{-k\downarrow} + \text{H.c.}).
\end{split}
\end{equation}

Here $n=\sum_\sigma d^\dag_\sigma d_\sigma$ is the localised level occupancy operator, $\delta$ its energy position (later we shall scale $\delta$ to $V_{G3}$ for a direct comparison with the experimental data) and $S_z=(d^\dag_\uparrow d_\uparrow-d^\dag_\downarrow d_\downarrow)/2$ its spin operator.
The coupling between the level and the leads, $v$, results in a broadening  $\Gamma=\pi v^2 \rho$, where $\rho=\sum_k \delta(\omega-\epsilon_k)$ is the density of states in the leads. 
The operator quantity $-x n_{d,\bar{\sigma}}$ corresponds to a correlated-hopping term introducing a perturbation of the level hybridization whose magnitude depends on its occupation for opposite spins $n_{d,\bar{\sigma}}$ \cite{Meir2002}. We find that $x$ = -0.4 yields the best agreement with the data. 
We assume $\Gamma$ to depend quadratically on $\delta$ (and hence on $V_{G3}$) and $B$ through the relation $\Gamma=\Gamma_0 + \Gamma_1 (c_0 + c_1 \delta/U + c_2
E_Z/U)^2$. The $B$ dependence can be expected from the influence of the magnetic field on the orbital motion and confinement of electrons. 
The last term in Eq. (\ref{hamiltonian}) accounts for superconducting pairing, with $\Delta$ being the induced superconducting gap. 

The model was solved using the numerical renormalization group (NRG)
method \cite{wilson1975,bulla2008}.

We begin by applying the model to the normal regime ($\Delta =0$) at high $B$. We find that the free 
parameters of the model are severely constrained even if only qualitative features
of the conductance are to be reproduced for different $T$ and $B$.  In this sense, the model is robust.
At $B=0$, Kondo correlations at finite $T$ enhance the conductance to a value below the unitary limit producing a 0.7-type conductance shoulder at $\delta$ $\sim 0$, as experimentally observed at $T = 4.2$ K (Fig. \ref{fig:image2}b).  At finite $B$, the shoulder evolves into a plateau at $0.9 e^2/h$. The results of the NRG calculations reproduce remarkably well the experimental trend as shown in Fig. \ref{fig:image1}d. 
In particular, the calculated conductance at the spin-resolved 0.7 anomaly remains constant despite the large variation of the $E_{z}/U$ ratio.
The $B$-dependent term (proportional to $c_2$), even if small against the gate dependent term (proportional to $c_1$), is essential to produce this behavior. Without it, the plateau would evolve into a local minimum, as we actually find experimentally when $B$ is applied perpendicularly to the NW under the same gate configuration (see Supplementary  Material).

\begin{figure}[!htb]
\includegraphics[width=86 mm]{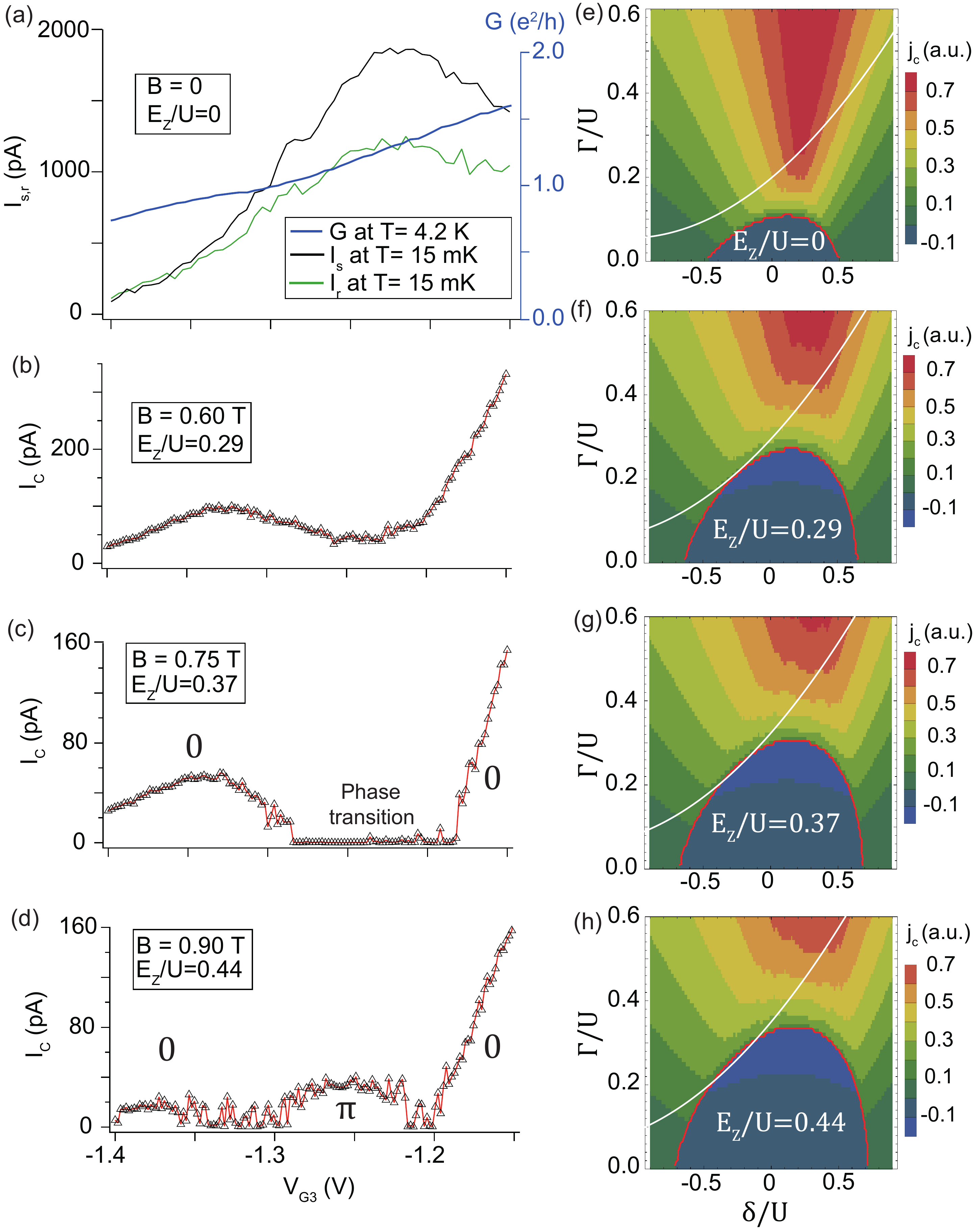}
\caption{
Left panels (a-d): Measured $V_{G3}$ dependence of the switching currents $I_r$ and $I_s$ at $B=0$ (a) and of the fitted critical current $j_c$ at $B=0.6$ T (b), $B=0.75$ T (c), and $B=0.9$ T (d). The normal-state $G$ measured at 4.2 K and $B=0$ is overlaid in panel (a). 
Right panels (e-h): NRG calculations of the $j_c$ as a function $\delta/U$ and $\Gamma/U$, for different values of $E_z/U$. Each phase diagram has a closed region corresponding to a spin-1/2 ground state, surrounded by an open region where the ground state is a singlet. Crossing the boundary between these regions at constant superconducting phase difference results in a reversal of the supercurrent.  To underline this effect, in the spin-1/2 region we conventionally give a negative sign to $j_c$ .  
The white lines represent the $\Gamma(\delta, E_Z)$ dependence obtained from the normal state fit parameters.
}
\label{fig:image3}
\end{figure}

We are now ready to address the superconducting proximity effect in the 0.7-anomaly regime.
Figs. \ref{fig:image3}a-d show supercurrent measurements as a function $V_{G3}$ at different values of $B$. Except for panel a, showing switching and re-trapping currents directly measured at $B=0$, the other panels display $j_c(V_{G3})$ traces obtained from fitting to the so-called resistively and capacitively shunted junction (RCSJ) model (details on the measurement and fitting methodology are given in the Supplementary Material). 
Remarkably, while the normal conductance increases monotonically with $V_{G3}$ (see the superimposed $G(V_{G3})$ trace in panel a), $j_c$ does not, in contrast to the Ambegaokar-Baratoff relation, for which $j_c  \sim G \Delta$. At $B=0$, the switching currents (closely related to $j_c$) are slightly peaked in correspondence with the 0.7-anomaly regime. Upon increasing $B$, $j_c(V_{G3})$ develops a minimum around $V_{G3} = - 1.25 V$ (panel g) and gets fully suppressed for $B = 0.75$ T (panel h) before re-emerging at higher $B$ (panel i). 

The above behavior can be explained using the Hamiltonian in Eq. (\ref{hamiltonian}).
Figs. \ref{fig:image3}e-h show NRG calculations of $j_c$, as a function of $\delta/U$ and $\Gamma/U$. We used $\Delta = 0.08$ meV, as deduced from 
tunnel spectroscopy measurements close
to full channel depletion (see Supplementary Material).
The plots depict phase diagrams consisting
of an open region where $j_c > 0$ (corresponding to a spin-singlet ground state) and a closed region where $j_c< 0$ (corresponding to a spin-1/2 ground state). 
The sign reversal reflects a $\pi$ phase shift in the current-phase Josephson relation.
  
Due to the positive sign of the $x$ parameter, the phase boundary has the shape of a skewed arc leaning to the right.
The $B=0$ case (Fig. \ref{fig:image3}e) has been extensively studied both theoretically \cite{Meng2009} and experimentally \cite{cleuziou2006carbon,van2006supercurrent,Jorgensen2007,Eichler2009,Maurand2012,delagrange2017,lee2017scaling}.
In the odd-charge regime ($-0.5 < \delta < 0.5$), strong (weak) coupling tends to stabilize a singlet (doublet) ground state. 
The singlet has a predominantly BCS character for $\Delta \gg \Gamma_S \gg U$, and a predominatly Kondo character for $\Gamma_S > \Delta$. The Zeeman effect contrasts both of these many-body phenomena thereby reducing the singlet binding energy and making the spin-1/2 domain grow (Figs. \ref{fig:image3}f-h) \cite{lee2014spin}. 

The above phase diagrams can account for the unusual, non-monotonic $B$ dependence of $j_c$ observed experimentally. The white lines in Figs. 3e-h denote the $\Gamma(\delta, E_Z)$ trajectory followed in the experimental sweeps, as deduced from normal-state fit parameters. As the doublet region of the phase diagram grows with $E_Z$, its phase boundary approaches the $\Gamma(\delta, E_Z)$ trajectory leading to a suppression of $j_c$ in the region of closest proximity. At $E_Z/U = 0.37$, the phase boundary reaches the $\Gamma(\delta,E_Z)$ trajectory and $j_c$ is correspondingly fully suppressed due to a competition between 0- and $\pi$-junction behavior. For larger $E_Z$, the $\Gamma(\delta, E_Z)$ trajectory crosses the spin-1/2 region within which the system acquires a clear $\pi$-junction behavior characterized by the emergence of a negative $j_c$.

In conclusion, we have shown that, even when seemingly absent, charge localization may play a crucial role in the transport properties of semiconductor NWs. The herein employed multi-gate device geometry proved to be essential towards clarifying this behavior. We have further shown that charge localization in a NW junction gives rise to a strong non-monotonic behavior of the Josephson current as a function of $B$ due to 0.7 physics. Our findings own relevance also in relation to  experiments aiming at detecting Majorana modes in Josephson junction geometries based on depleted NWs under strong Zeeman fields \cite{tiira17,0957-4484-28-8-085202,zuo2017supercurrent}. In particular, the predictions based on the anomalous B-field dependence of the critical current owing to the presence of Majoranas in the junction \cite{PhysRevLett.112.137001,PhysRevB.96.205425,san2013multiple} may be masked by the localization effects and the 0.7 physics discussed here.

We acknowledge financial support from the Agence Nationale de la Recherche, through the TOPONANO project, and from the EU through the ERC grant No. 280043. R. A. acknowledges financial support from the Spanish Ministry of Economy and Competitiveness through Grant FIS2015-64654-P. 
R.~\v{Z}. acknowledges the support of the Slovenian Research Agency (ARRS) under Program P1-0044 and J1-7259.

%

\pagebreak
\widetext
\begin{center}
\textbf{\large Supplementary Materials: Supercurrent through a spin-split quasi-ballistic point contact in an InAs nanowire}
\end{center}
\setcounter{equation}{0}
\setcounter{figure}{0}
\setcounter{table}{0}
\setcounter{page}{1}
\makeatletter
\renewcommand{\theequation}{S\arabic{equation}}
\renewcommand{\thefigure}{S\arabic{figure}}
\renewcommand{\bibnumfmt}[1]{[S#1]}
\renewcommand{\citenumfont}[1]{S#1}

\section{Lever-arm parameter}

\begin{figure}
	\includegraphics[width=0.5\linewidth]{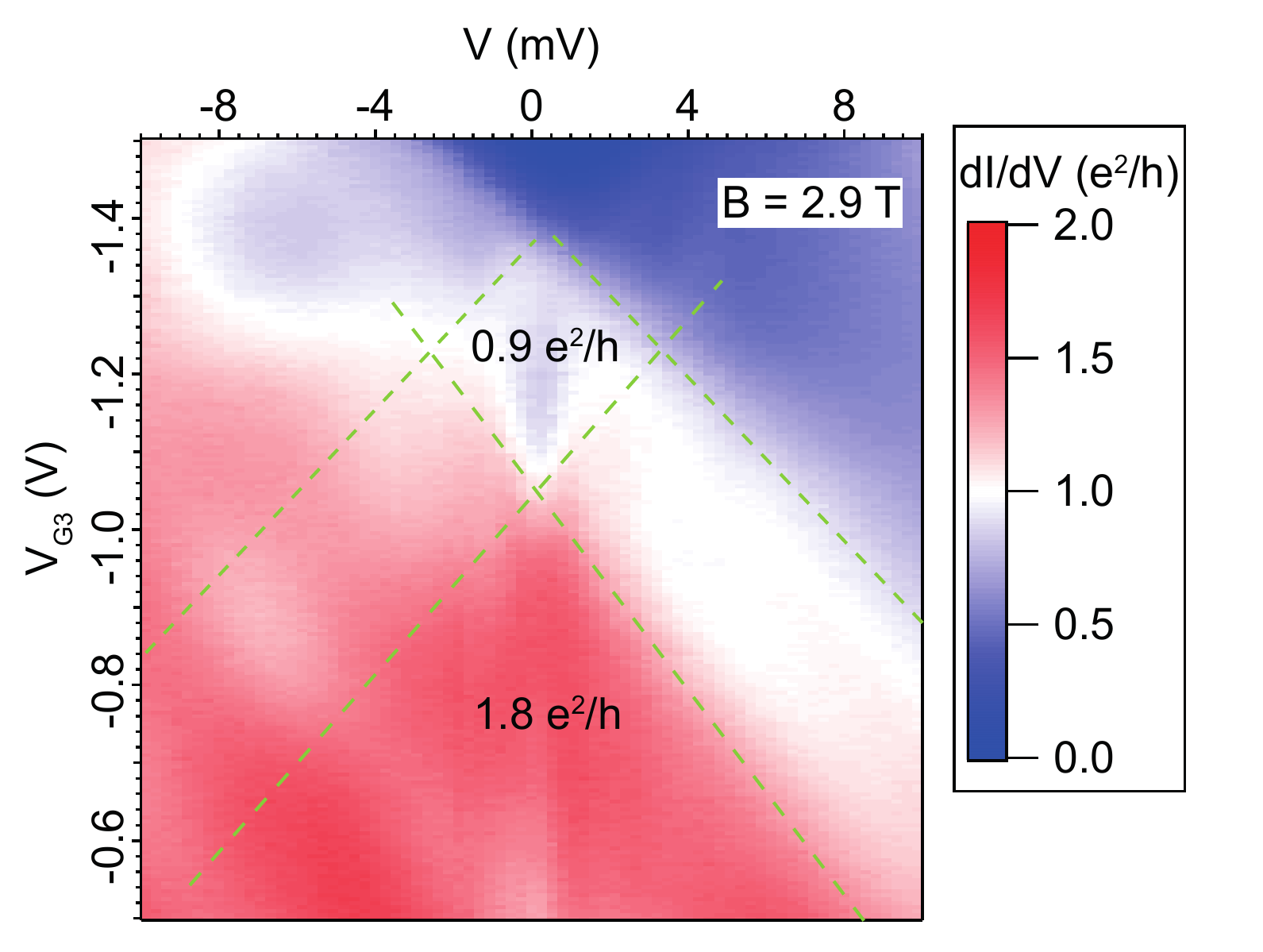} 
\caption{Differential conductance as a function of bias voltage and $V_{G3}$, taken at $B = 2.9$ T. The lever-arm parameter was extracted from the slope of the green-dashed lines, which correspond to high-bias features of the spin-split channel.}
\label{fig:Simage1}
\end{figure}

The lever-arm parameter $\alpha $ = 0.0082 meV/mV used to convert $V_{G3}$ to energy was extracted from the differential conductance (dI/dV) colormap taken at $B = 2.9$ T shown in Fig. \ref{fig:Simage1}.

\section{Resonances before the conductance plateau}

The linear conductance exhibited a few quantum dot resonances at $V_{G3}$ more negative than the region displaying quantized conductance at large field shown in the main text. These resonances are plotted in Fig. \ref{fig:Simage11}a, together with the parity of the states (E:even; O:odd). The width of these resonances increases with $V_{G3}$, corroborating our assumption of a tunnel coupling dependent on this gate. This succession of Coulomb resonances confirms the odd parity of the state whose screening results in the 0.7 anomaly. 
    
\section{Many-subbands regime}

At certain gate configurations, the two barriers of the quantum dot described in the main text could be best described as two quantum point contacts (QPCs) in series. We call them QPC 2 and QPC 3, for their location in the sections of the nanowire above gates 2 and 3, respectively (see the scheme in Fig. \ref{fig:Simage4}a). 

\begin{figure}
	\includegraphics[width=0.6\linewidth]{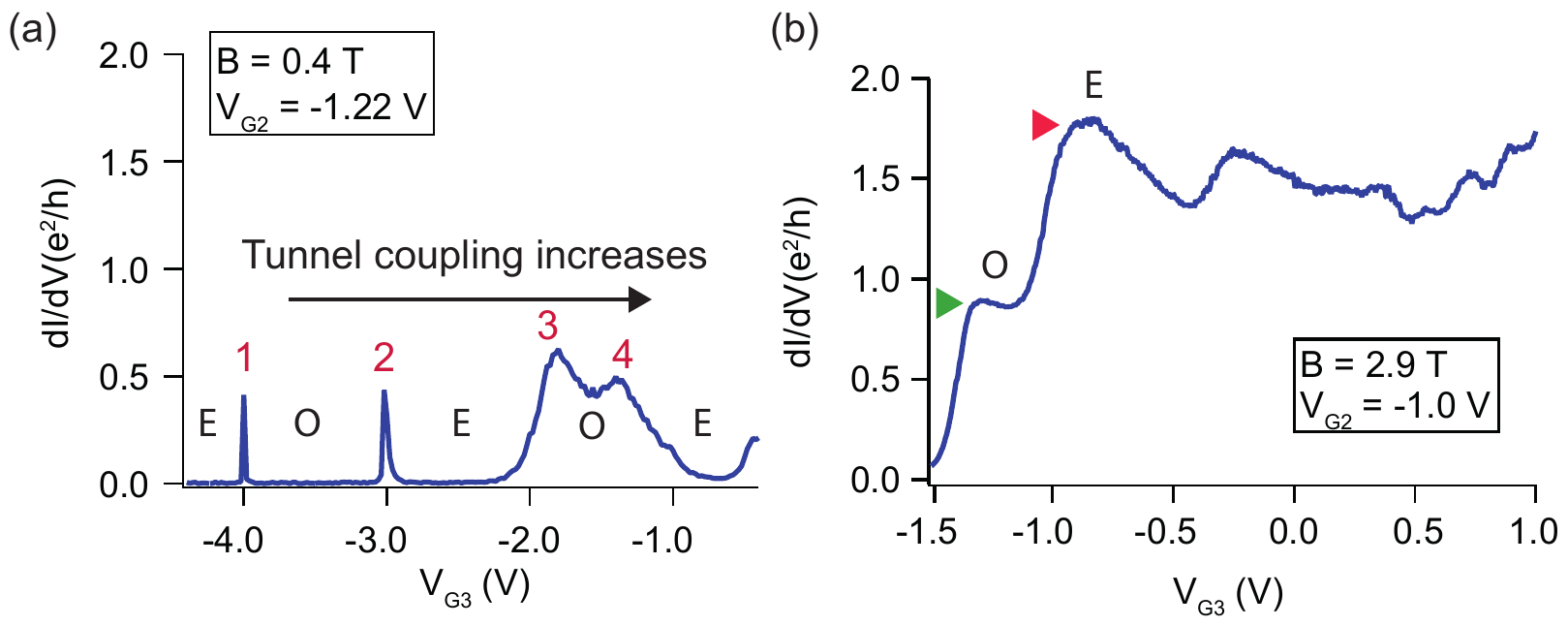} 
\caption{Linear conductance at $V_{G2}$ =-1.22 V, showing an example of the resonances preceding the 0.7 anomaly. E (O) corresponds to an even (odd) number of electrons in the quantum dot.}
\label{fig:Simage11}
\end{figure}

\begin{figure}
	\includegraphics[width=0.5\linewidth]{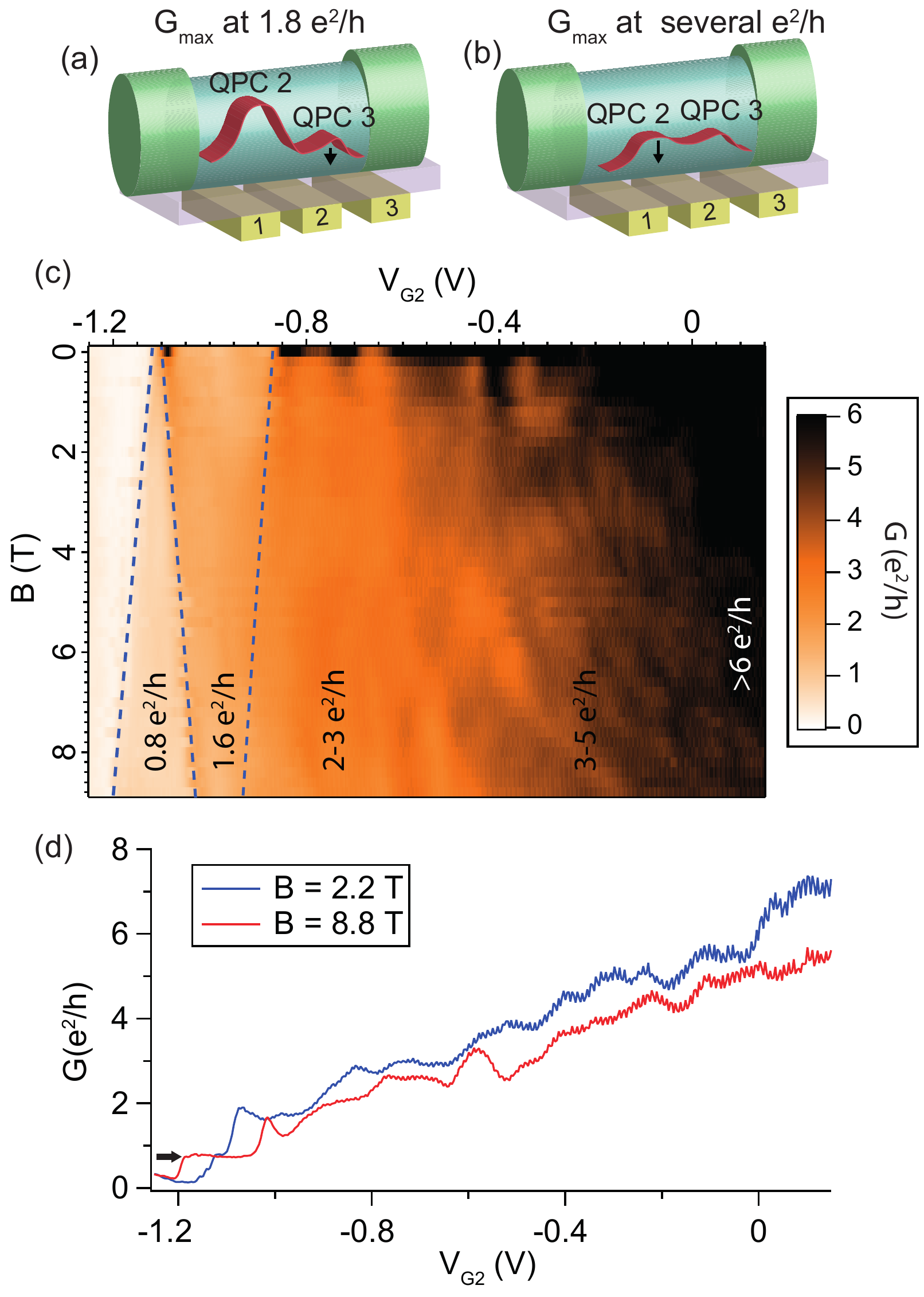}
\caption{(a,b) Schemes of the conduction band profile of along the channel of the nanowire. Two quantum point contacts in series may form in the channel of the nanowire above gates 2 and 3. A black arrow shows the way the gates were swept in (a) Figs. 2 and 3 of the main text, and in (b) this section. The maximum conductance attained in each case is also indicated. (c) Colormap of linear conductance vs. magnetic field and $V_{G2}$ taken at $V_{G3}= 1 V$ and $T=15 mK$. The conductance of the first two plateaus is indicated, and their Zeeman splitting followed by blue-dashed lines. (d) Cuts at $B=2.2\si\tesla$ and $B=8.8\si\tesla$, above the critical field of the superconductor. A plateau at $0.8\si{\elementarycharge}^{2}/$h$ $ whose width grows with field is pointed by a black arrow.}
\label{fig:Simage4}
\end{figure}

In the main text, QPC 2 was kept nearly closed at a negative voltage of $V_{G2}=-1 V$, which restricted the conductance to values below $2\si{\elementarycharge}^{2}/$h$ $, no matter how much we opened QPC 3 by pushing $V_{G3}$ to positive voltage (see Fig. \ref{fig:Simage11}b). QPC 2 enforced in this case a one-subband regime.

The ample tunability of our device also allowed us to explore a regime of transport through many subbands, in which case QPC 3 was open and QPC 2 was varied (see the scheme in Fig. \ref{fig:Simage4}b). In this new gate configuration, gate 3 was fixed at $V_{G3} =1 V$, and $V_{G2}$ was swept.  

Figure \ref{fig:Simage4}c shows a measurement of the magnetic field evolution of G in this new configuration as a function of $V_{G2}$, with the field oriented at 45$^{\circ}$ with respect to the axis of the nanowire. In this plot, blue dashed lines were added to follow the Zeeman splitting of conductance plateaus. Two cuts taken from this plot at $B>B_{c}$ (displayed in Figure \ref{fig:Simage4}d) show that a clear conductance plateau develops at $0.8\si{\elementarycharge}^{2}/$h$ $, denoting that a QPC regime exists at this gate voltage.

These measurements, together with the ones of the main text, provide a global picture of the way the conduction-band  profile of the nanowire can be altered by the two gates, leading either to low-transparency localization, QPC conductance quantization or, remarkably, to a localized magnetic impurity that mimics conductance quantization at large field. 

\section{Dependence of the linear conductance on the magnetic field when aligned perpendicular to the axis of the nanowire}
\label{sec:perpendicularfield}

\begin{figure}
	\includegraphics[width=0.85\linewidth]{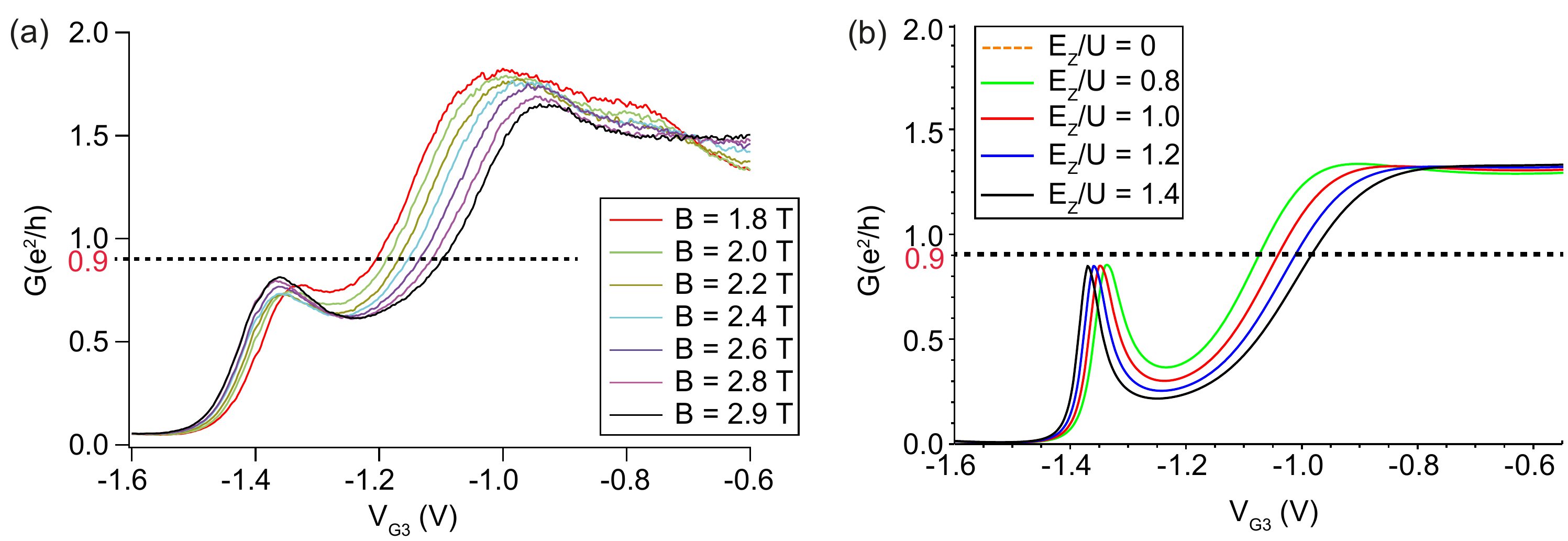} 
\caption{(a) Magnetic field dependence of the zero-bias conductance in the gate region of the 0.7 anomaly when the field is applied perpendicularly to the axis of the nanowire. Since the g-factor in this direction is larger (g = 11.5) than in the parallel direction, the $E_{z}$/U ratio covers a larger range: $E_{z}/U$ = 0.76 (for 1.8 T) to 1.35 (for 2.9 T). (b) NRG calculation of the conductance for the same range of $E_{z}$/U as the measurements in (a), with a model similar to the calculation for parallel magnetic field as described in the main text, but without the ($c_2$) term.}
\label{fig:Simage5}
\end{figure}

Figure \ref{fig:Simage5}a shows the perpendicular magnetic field evolution of G at a field large enough to suppress superconductivity. Instead of a plateau of quantized conductance -as observed for parallel magnetic field under the same conditions (main text)-, there is a peak followed by a dip in the conductance. Furthermore, the conductance in the dip \textit{decreases} as the magnetic field is increased. 

This magnetic field behavior can be qualitatively understood if we assume that the tunnel coupling $\Gamma$ does not increase with the magnetic field (but just with $V_{G3}$), which results in a more pronounced Coulomb blockade effect. Figure \ref{fig:Simage5}b shows that, indeed, the NRG model as described in the main text can qualitatively replicate the experimental data of Fig. \ref{fig:Simage5}a when, as opposed to the parallel-field case, we neglect the quadratic term arising from the $B$-contribution by setting $c_2=0$ in $\Gamma$ ($V_{G3}$,$B$).

The observation of a conductance plateau for parallel field, and Coulomb blockade oscillations for perpendicular field, confirms once more to the quantum dot nature of the 0.7 anomaly in the device. This is distinctively different from a conventional QPC behavior, where conductance quantization occurs regardless of the field direction, aside from backscattering suppression. 

\section{Equivalent circuit and quality factor of the Josephson junction}

The nanowire Josephson junction was underdamped at low magnetic field and/or with low critical current (which happened at low normal-state conductance), and mostly overdamped at high magnetic field. In this section we explain why this is the case.

\begin{figure}
	\includegraphics[width=1\linewidth]{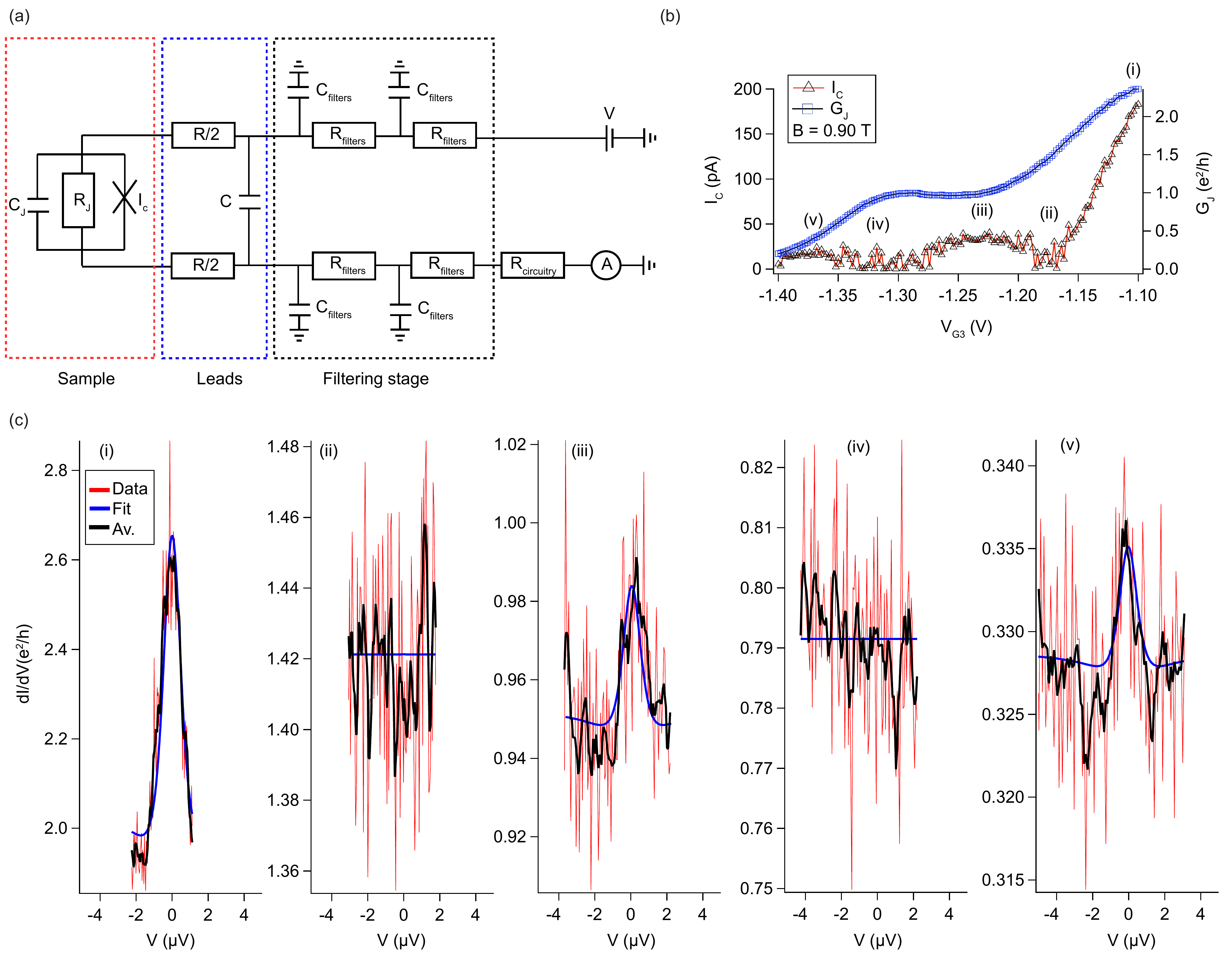}
\caption{(a) Schematic diagram of the equivalent circuit of the measurement setup. (b) Fitted critical current ($I_c$) and junction conductance ($G_J$) as a function of gate 3 voltage, taken from corrected V - dI/dV data at B = 0.9 T. (c) Traces V - dI/dV of data (red line) that were fitted (blue line) to extract the $I_c$ and $G_J$ at the gate voltage points (i) to (v) indicated in (b). An eleven-point average of the data is shown as an aid to the eye (black line), to help identifying the small zero-bias peak in the data when $I_c$ is small.}
\label{fig:Simage6}
\end{figure}

Figure \ref{fig:Simage6}a shows the equivalent circuit of the Josephson junction device. The cold parts of the circuit, which were kept at T$=$15 mK during the measurement, are encircled by dashed lines. The voltage source and the ammeter are at room temperature; the latter in series with a resistance $R_{circuitry}$ of about 10 k$\Omega$ from the current amplifier. 

The cold parts of the circuit consisted of a filtering stage (black dashed lines), the on-chip leads (blue dashed lines), and the sample (red dashed lines). The filtering stage had a two-stage RC circuit, with $R_{filters}$=10 k$\Omega$ and $C_{filters}$=10 nF. Since there are four of these resistances in series with the sample, the total resistance $R_s$ in series with the sample was of about 50 k$\Omega$.

The on-chip leads capacitance was estimated at $C=1\times10^{-15}$ F from the capacitance of two neighboring bonding pads. The resistance $R$ of the leads was determined by fitting of the supercurrent V-$dI/dV$ characteristic to be R$=$1.6 k$\Omega$, as explained in the next section.

In the circuit of Figure \ref{fig:Simage6}a, the sample itself is modeled as a Josephson supercurrent source I($\phi$) of critical current $I_c$, in parallel with a junction resistance $R_J$ and a junction capacitance $C_J$. $I_c$ and $R_J$ depended on the gate voltage and ranged, respectively, from 20 pA to 2 nA, and from 20 k$\Omega$ to 100 k$\Omega$. $R_J$ could also depend on the bias voltage $V$, but for the small bias applied on the device at high magnetic field -of less than 20 $\mu V$-, this dependency could be dropped. The value of $C_J$ could be estimated from the charging energy U$=$1.3 meV, from which we obtain $C_J=6\times10^{-17}$ F.

In the resistively and capacitively shunted junction (RCSJ) model, the quality factor $Q$ of the junction can be evaluated by the following formula \cite{S_J??rgensen2007}: 

\begin{equation} \label{Quality}
Q=\frac{\sqrt{\hbar[C(1+R/R_J)+C_J]/(2eI_c)}}{RC+\hbar/(2eI_cR_J)}
\end{equation}

For our sample, the $Q$ factor is gate-dependent because $R_J$ and $I_c$ are not inversely proportional to each other for all gate voltages. Since $I_c$ and $R_J$ are also changing with magnetic field, $Q$ will be a function of the gate voltage \textit{and} the magnetic field.

Nevertheless, it is possible to roughly estimate $Q$ for a few values of $R_J$ and $I_c$ and see its tendency. At $B= 0$, $Q = 1.5$ at the supercurrent maximum of Figure 3a of the main article and therefore the junction is underdamped for this \textit{particular} gate voltage. At the supercurrent minimum of the same plot, when the normal conductance of the sample is low, $Q = 0.8$ and the junction is slightly overdamped. At this gate voltage, $I_s$ and $I_r$ are equal. Since $I_c$ tends to decrease with a rising magnetic field, $Q$ becomes smaller as the magnetic field increases, and the junction is predominantly \textit{overdamped}.

\begin{figure}
	\includegraphics[width=0.9\linewidth]{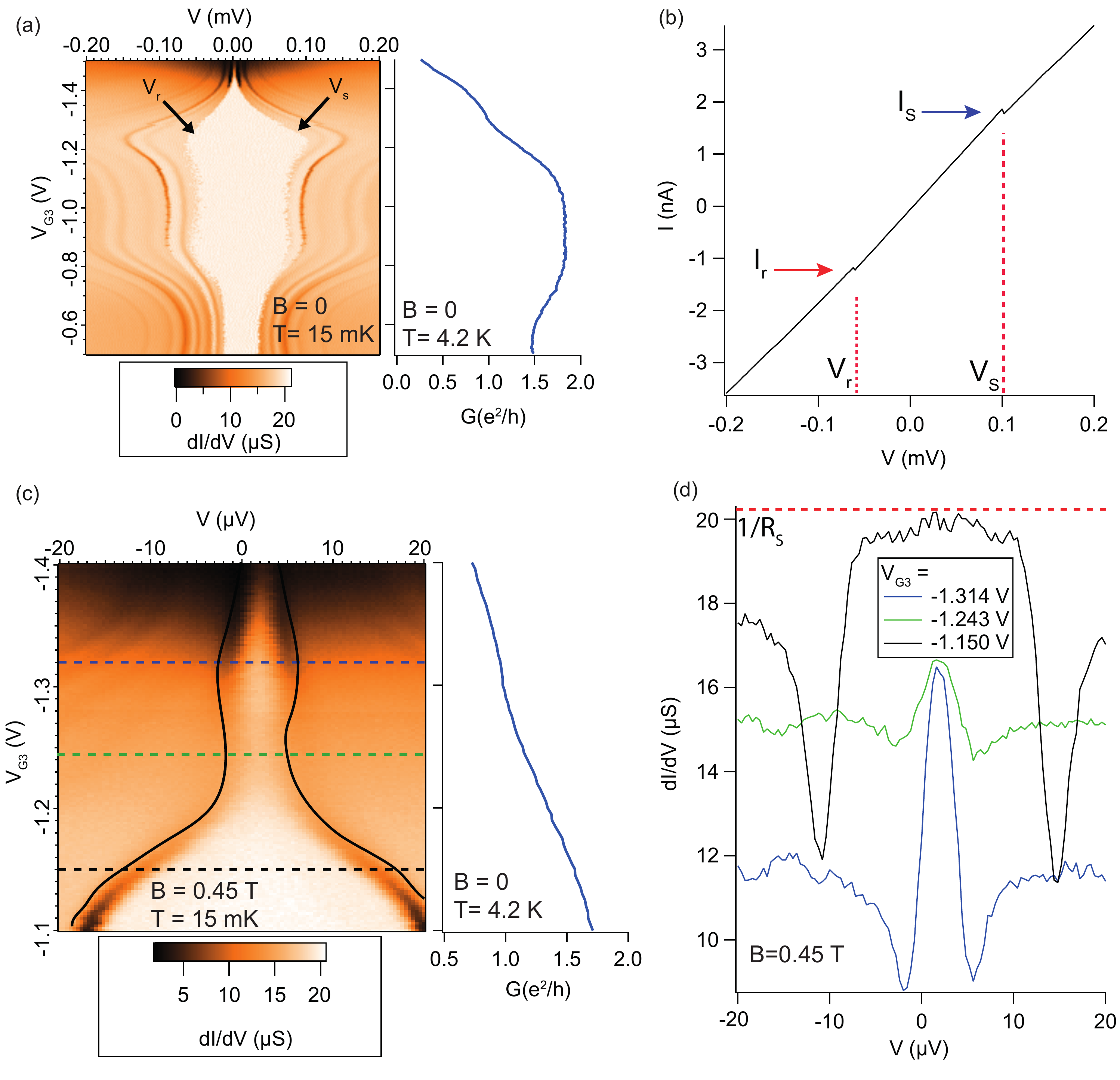} 
\caption{(a) Raw-data conductance map taken at B = 0. A non-dissipative supercurrent appears as a plateau of conductance $1/R_s$ around zero-bias. The quantities $V_r$ and $V_s$, which are proportional to $I_r$ and $I_s$, are indicated by black arrows. The corresponding normal-state linear conductance (Figure 2b of the main article) is presented in the inset on the right of the map. (b) Trace of raw V-I data taken at the maximum of supercurrent, in the middle of the 0.7 anomaly. Since $|I_s|>|I_r|$, the junction is underdamped at this field. (c) Raw-data conductance map taken at B = 0.45 T, with black lines added to roughly follow $V_r$ and $V_s$ as the gate is swept. The corresponding normal-state linear conductance (Figure 2b of the main article) was added in the inset on the right of the map. If one corrects a voltage offset of 2 $\mu$V from the source, then $|V_r|=|V_s|$, indicating that the junction is overdamped at this field. (d) Raw conductance traces taken from the dashed lines in (c) of the corresponding color. The zero-resistance state -i.e., when the conductance $1/R_s$ is reached-, is indicated by a red-dashed line.}
\label{fig:Simage9}
\end{figure}

Figures \ref{fig:Simage9}a,c show raw-data conductance maps as a function of the gate voltage $V_{G3}$, at $B = 0$ and at $B = 0.45$ T, respectively. In these maps, whenever the supercurrent is non-dissipative, it appears as a plateau of conductance $1/R_s$ around zero-bias, such as in the black curve in Figure \ref{fig:Simage9}d. This plateau, if the junction is underdamped, will be bound by $V_r$ and $V_s$, which are proportional to the re-trapping ($I_r$) and switching ($I_s$) current. $V_r$ and $V_s$ are indicated in the map of Figure \ref{fig:Simage9}a. $|V_s|>|V_r|$ in most of this gate range, as for an underdamped junction. To extract $I_s$ and $I_r$, we took V-I curves like the one in Figure \ref{fig:Simage9}b, in which these quantities are indicated. In this curve, $I_s$ and $I_r$ are easily distinguishable.

At B = 0.45 T, the junction becomes overdamped for all the gate range shown in the raw data of Figure \ref{fig:Simage9}c. This is revealed by the symmetry of the supercurrent with respect to zero bias, if one corrects for a 2 $\mu$V voltage offset from the voltage source. The magnetic field renders the junction \textit{overdamped}, consistently with our estimation of Q.

Figure \ref{fig:Simage9}d shows three raw conductance traces taken from the map (along the dashed lines of the same color as the corresponding curves). At this magnetic field, the supercurrent is non-dissipative only around the black dashed line, as shown in the black curve, whose conductance approaches $1/R_s$. At lower gate voltage, there is a zero-bias peak instead of a plateau, and its conductance is clearly below $1/R_s$. Here the supercurrent manifests itself as a dissipative conductance peak, which occurs because the Josephson energy becomes so small that it attains the same order of magnitude as the thermal energy. In the next section, we detail how the critical current was extracted in this case.

\section{Method for fitting the supercurrent}
\label{sec:SupercurrentFit}

The critical current was obtained from a fit of the corrected V - $dI/dV$ data with the theory derived from the RCSJ model with thermal noise in Ref. \cite{S_Anchenko1969}, extended in Ref. \cite{S_Halperin} and used in Refs. \cite{S_J??rgensen2007,S_Eichler2009,steinbach2001direct}. The correction of the voltage and the conductance consisted in subtracting the series resistance $R_s$ according to: $V=V_{raw}-IR_s$ and $G=G_{raw}/(1-R_{s}G_{raw})$. We assumed, as also done in Refs. \cite{S_J??rgensen2007,S_Eichler2009}, that the current-phase relationship was sinusoidal ($I_{s}=I_{c}sin(\phi)$). This assumption may no longer hold near the singlet $\to$ doublet transition point \cite{PhysRevB.93.195437}. Since we took V - $dI/dV$ measurements, the derivative $dI/dV$ of the original formula given in equation \ref{Langevin} was taken, where $I_{\alpha}(x)$ is the modified Bessel function of complex order $\alpha$, $\eta(V)={\hbar}V/2eR\kappa_{B}T$, and $\beta=I_{c}{\hbar}/2e\kappa_{B}T$ is the ratio between the Josephson energy and the thermal energy. This equation contains the resistance of the junction $R_J$, which was added to the original expression given in Ref. \cite{S_Anchenko1969} to account for an additional multiple Andreev reflection (MAR) channel in parallel with the Josephson current \cite{S_J??rgensen2007}. The resistance $R_J$ provides an ohmic contribution at a current above the critical current.

\begin{equation} \label{Langevin}
I(V)=\frac{R_{J}}{R_{J}+R}\left(I_cIm\left[\frac{I_{1-i\eta(V)}(\beta)}{I_{-i\eta(V)}(\beta)}\right]+\frac{V}{R_{J}}\right)
\end{equation}

For a small Josephson energy with respect to the thermal energy (i.e., $\beta\ll1$), which was the case for $B\geq0.6\si\tesla$, the dI/dV expression simplifies to \cite{S_Anchenko1969}:

\begin{equation} \label{fit2}
\frac{dI}{dV}(V)=\frac{R_J}{R_J+R}\left[\frac{I_c^2R\eta^2\left(1-\eta^2\right)}{2V^2\left(1+\eta^2\right)^2}+\frac{1}{R_J}\right]
\end{equation}

Both equations give similar results for $B\geq$0.6 T. An example of five fits (i-v) that produce the $I_c$ and $G_J=1/R_J$ data points indicated in the re-emergent supercurrent plot of Figure \ref{fig:Simage6}b is given in Figure \ref{fig:Simage6}c. As it is seen in this series of plots, the supercurrent arises as a narrow -a few \si{\micro\eV} wide- and small zero-bias peak in the conductance. A large supercurrent produces a large zero-bias peak (i) -and viceversa (iii and v). When there is no peak (ii and iv), as it occurs in a $\phi_{0}=0$ $\to$ $\phi_{0}=\pi$ phase transition, the critical current is zero or attains a very small value below 20 pA. We fit the data at all magnetic fields studied with $T= 15 mK$ and $R = 1.6$ k$\Omega$ as fitting parameters.

\section{Superconducting gap}

\begin{figure}
	\includegraphics[width=0.5\linewidth]{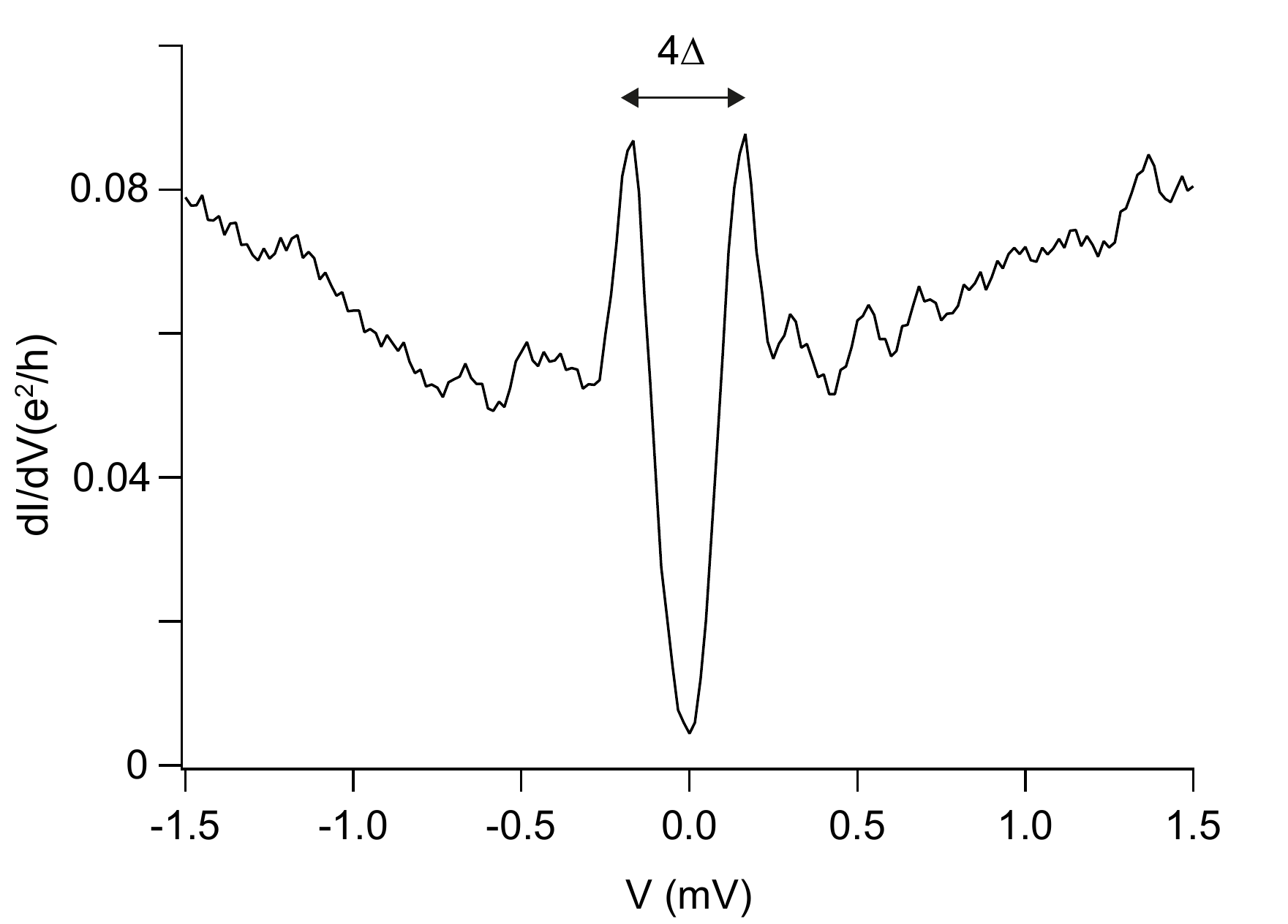}
\caption{Measurement of the superconducting gap, at $B = 0$, $T = 15 mK$, $V_{G3} = -1.6 V$, $V_{G2}= -1 V$. The bias separation between the BCS peaks is $4\Delta = 0.32$ meV.}
\label{fig:Simage10}
\end{figure}

Fig. \ref{fig:Simage10} shows a plot of the superconducting gap with the nanowire near depletion, in the tunnel regime. $\Delta = 0.08$ meV is extracted from this measurement, in agreement with the $T_c$ of thin Ta films evaporated with a similar procedure as the one used for the contacts of the device. This value was used in the NRG calculations of the supercurrent through the spin-split single level. 

\section{NRG calculations}

The numerical renormalization group (NRG) calculations for the
Hamiltonian [Eq.~(1) in the main text] have been performed to determine
the normal-state differential conductance (for $\Delta=0$) and the
Josephson current (for $\Delta\neq0$) in the presence of the magnetic
field. For normal-state properties, we used NRG discretization
parameter $\Lambda=2$, two interleaved discretization meshes, keeping
up to 5000 multiplets (or using an energy cutoff at energy $10$ in the
units of the characteristic energy scale of a given NRG step). The
conductance was extracted from raw dynamical properties data without
performing a spectral broadening. The calculations in the
superconducting state were performed with $\Lambda=8$, keeping up to
10000 multiplets (or using an energy cutoff at energy $6$). These
calculations were performed for a range of phase difference $\phi$,
from which we extracted the critical current defined as the Josephson
current $j(\phi)$ such that the absolute value $|j(\phi)|$ is
maximized.

\medskip

\end{document}